\documentclass[aps, prd, floatfix, twocolumn, superscriptaddress, nofootinbib]{revtex4-1}

\usepackage{latexsym}
\usepackage{amsmath}
\usepackage{amssymb}
\usepackage{amsfonts}

\usepackage{dsfont}
\usepackage{slashed}
\usepackage{bm}
\usepackage{urwchancal}
\usepackage{ulem}
\usepackage[vcentermath]{youngtab}
\usepackage{color}
\definecolor{purple}{rgb}{0.5,0,0.5}
\definecolor{blue}{rgb}{0.0,0,0.9}
\definecolor{prdblue}{rgb}{0.133,0.118,0.498}
\usepackage[colorlinks=true, pdfstartview=FitV, linkcolor=prdblue, citecolor= prdblue, urlcolor=prdblue]{hyperref}

\usepackage{enumitem}
\usepackage{multirow}%for tabular
\usepackage{subfigure}
\usepackage{textcomp}%for +-

\usepackage{supertabular} 
\usepackage{placeins}
\usepackage{epsfig}
\usepackage{graphicx}

\def\tstrut{\vrule height3.25ex depth0pt width0pt} % used in tables

\begin{document}

%\preprint{}

\title{Towards the discovery of novel $B_c$ states: radiative and hadronic transitions}

\author{B. Mart\'in-Gonz\'alez}
\email[]{id00697965@usal.es}
\affiliation{Departamento de Física Fundamental, Universidad de Salamanca, E-37008 Salamanca, Spain}

\author{P. G. Ortega}
\email[]{pgortega@usal.es}
\affiliation{Departamento de Física Fundamental, Universidad de Salamanca, E-37008 Salamanca, Spain}
\affiliation{Instituto Universitario de F\'isica Fundamental y Matem\'aticas (IUFFyM), Universidad de Salamanca, E-37008 Salamanca, Spain}

\author{D. R. Entem}
\email[]{entem@usal.es}
\affiliation{Instituto Universitario de F\'isica Fundamental y Matem\'aticas (IUFFyM), Universidad de Salamanca, E-37008 Salamanca, Spain}
\affiliation{Grupo de F\'isica Nuclear, Universidad de Salamanca, E-37008 Salamanca, Spain}

\author{F. Fern\'andez}
\email[]{fdz@usal.es}
\affiliation{Instituto Universitario de F\'isica Fundamental y Matem\'aticas (IUFFyM), Universidad de Salamanca, E-37008 Salamanca, Spain}
\affiliation{Grupo de F\'isica Nuclear, Universidad de Salamanca, E-37008 Salamanca, Spain}

\author{J. Segovia}
\email[]{jsegovia@upo.es}
\affiliation{Departamento de Sistemas F\'isicos, Qu\'imicos y Naturales, Universidad Pablo de Olavide, E-41013 Sevilla, Spain}

\date{\today}

\begin{abstract}
The properties of the $B_c$-meson family ($c\bar b$) are still not well determined experimentally because the specific mechanisms of formation and decay remain poorly understood.
Unlike heavy quarkonia, \textit{i.e.} the hidden heavy quark-antiquark sectors of charmonium ($c\bar c$) and bottomonium ($b\bar b$), the $B_c$-mesons cannot annihilate into gluons and they are, consequently, more stable. The excited $B_c$ states, lying below the lowest strong-decay $BD$-threshold, can only undergo through radiative decays and hadronic transitions to the $B_c$ ground state, which then decays weakly. As a result of this, a rich spectrum of narrow excited states below the $BD$-threshold appear, whose total widths are two orders of magnitude smaller than those of the excited levels of charmonium and bottomonium.
In a different article, we determined bottom-charmed meson masses using a non-relativistic constituent quark model which has been applied to a wide range of hadron physical observables, and thus the model parameters are completely constrained. Herein, continuing to our study of the $B_c$ sector, we calculate the relevant radiative decay widths and hadronic transition rates between $c\bar b$ states which are below $BD$-threshold. 
This shall provide the most promising signals for discovering excited $B_c$ states that are below the lowest strong-decay $BD$-threshold. Finally, our results are compared with other models to measure the reliability of the predictions and point out differences. 
\end{abstract}

% insert suggested PACS numbers in braces on next line \phi
%\pacs{}

%\keywords{}

\maketitle

%%%%%%%%%%%%%%%%%%%%%%%%%%%%%%%%%%%%%%%%%%%%%%%%%%%%%%%%%%%%%%%%%%%%%%%%%%%%%%%%
%%%%%%%%%%%%%%%%%%%%%%%%%%%%%%%%%%%%%%%%%%%%%%%%%%%%%%%%%%%%%%%%%%%%%%%%%%%%%%%%

\section{Introduction}
\label{sec:Introduction}

The feasibility of studying experimentally the family of $c\bar b$ mesons was demonstrated by the CDF Collaboration at the Tevatron collider in $1998$~\cite{Abe:1998wi, Abe:1998fb} with the observation of the $B_c(1^1S_0)$ bound state.\footnote{The spectroscopic notation $n^{2S+1}L_J$ is used, where $n=1$ indicates the ground state and $n=2,\,3,\ldots$, the respective excited states with higher energies but equal $J^P$ (following the notation of PDG), the total spin of the two valence quarks is denoted by $S$, while $L$ is their relative angular momentum where $S,\,P,\,D,\,F\,\ldots$ implies, respectively, $L=0,\,1,\,2,\,3,\ldots$, and $J$ is the total angular momentum of the system.} However, low production cross-sections, large backgrounds and relatively-easy misidentifications eluded the discovery of new bottom-charmed mesons until $2014$, when the ATLAS Collaboration~\cite{Aad:2014laa} observed a peak at $6842\pm4\pm5\,\text{MeV/c}^2$, which was interpreted as either the $B_c^\ast(2^3S_1)$ excited state or an unresolved pair of peaks from the decays $B_c(2^1S_0)\to B_c(1^1S_0)\pi^+\pi^-$ and $B_c^\ast(2^3S_1)\to B_c^\ast(1^3S_1)\pi^+\pi^-$ followed by $B_c(1^3S_1)\to B_c(1^1S_0)\gamma$. It was not until 2019 when the CMS~\cite{Sirunyan:2019osb} and LHCb~\cite{Aaij:2019ldo} Collaborations released signals consistent with the $B_c(2S)$ and $B_c^\ast(2S)$ states, observed in the $B_c(1S)\pi^+\pi^-$ invariant mass spectrum. More results on $B_c$ mesons are expected to be reported in the near future.

On the theoretical side, the $B_c$-meson family provides another opportunity to test non-relativistic quark models that have been successfully applied to charmonium ($c\bar c$) and bottomonium ($b\bar b$) systems. This is because the $B_c$ states share dynamical properties with both the $c\bar c$ and $b\bar b$ sectors, but they consist on two heavy quarks with different flavors that makes the $B_c$ states very stable, with narrow widths, since annihilation into gluons is forbidden. In fact, their results~\cite{Kwong:1990am, Eichten:1994gt, Eichten:2019gig, Fulcher:1998ka, Li:2019tbn, Tang:2022xtx} can be contrasted with those from other theoretical frameworks such as relativistic quark models~\cite{Godfrey:1985xj, Zeng:1994vj, Gupta:1995ps, Ebert:2002pp, Ikhdair:2003ry, Godfrey:2004ya, Ikhdair:2004hg, Ikhdair:2004tj}, QCD sum rules~\cite{Wang:2012kw, Wang:2013cha}, continuum functional methods for QCD~\cite{Chang:2019wpt, Chen:2020ecu, Yin:2019bxe}, effective field theories~\cite{Brambilla:2000db, Penin:2004xi, Peset:2018ria, Peset:2018jkf}, and lattice QCD~\cite{Allison:2004be, Mathur:2018epb, Dowdall:2012ab}. A collection of all these results should provide a reliable template from which to compare the future experimental findings. In fact, there is some agreement about which conventional $B_c$ states must exist below the lowest strong-decay $BD$ threshold. There should be two sets of $S$-wave states, the $1P$ multiplet and some or all of the $2P$ states, one multiplet of $D$-wave nature, and the lowest $F$-wave case should be located so close to threshold that its member states may be narrow due to angular momentum barrier suppression of the Okubo-Zweig-Iizuka (OZI)-rule~\cite{Okubo:1963fa, zweigcern2, iizuka1966systematics}.

Complications with the $B_c$ spectroscopy are expected to begin at the energy region in which strong-decay meson-meson thresholds could play an important role in the formation of $B_c$(-like) structures. This has been vigorously manifested in the heavy quarkonium spectrum with the discovery of many charmonium- and bottomonium-like $XYZ$ states~\cite{Brambilla:2010cs, Brambilla:2014jmp, Olsen:2014qna}. In Ref.~\cite{Ortega:2020uvc}, we studied the influence of two-meson thresholds on the $B_c$ states finding, for instance, dynamically generated additional states in the $J^P=1^+$ and $2^+$ channels very close to the $DB^\ast$ and $D^\ast B^\ast$ thresholds, respectively. In such article, however, we did not perform any study related with decay properties and possible ways of finding low-lying states located either below or around the lowest strong-decay meson-meson thresholds.

The theoretical methods used to study the spectroscopy of bottom-charmed mesons can be extended to their decay properties. The excited $B_c$ states lying below the $BD$-threshold can only undergo through radiative decays and hadronic transitions to the $B_c$ ground state, which then decays weakly. Therefore, radiative and hadronic decay rates almost comprise the total decay width of the lowest excited $B_c$ states, making them narrow with total widths two orders of magnitude smaller than those of the excited levels of charmonium and bottomonium, for which annihilation channels are significant. Moreover, such electromagnetic and hadronic processes are interesting by themselves because they allow experimental access to excited levels of heavy quarkonia which are below the lowest strong-decay meson-meson threshold and  provide information about the internal structure and quantum numbers.

In this article we extend our previous investigation of the $B_c$ spectrum~\cite{Ortega:2020uvc} to potentially interesting radiative decays and hadronic transitions. Our theoretical framework is a non-relativistic constituent quark model~\cite{Vijande:2004he} in which quark-antiquark and meson-meson degrees of freedom can be incorporated at the same time (see references~\cite{Segovia:2013wma} and~\cite{Ortega:2012rs} for reviews). The naive model, and its successive improvements, has been successfully applied to the charmonium and bottomonium sectors, studying their spectra~\cite{Segovia:2008zz, Segovia:2010zzb, Segovia:2015dia, Segovia:2016xqb, Yang:2019lsg}, their electromagnetic, weak and strong decays and reactions~\cite{Segovia:2009zz, vanBeveren:2010mg, Segovia:2011zza, Segovia:2012cd, Segovia:2013kg}, their coupling with meson-meson thresholds~\cite{Ortega:2009hj, Ortega:2016hde, Ortega:2017qmg, Ortega:2018cnm, Ortega:2021xst} and, lately, phenomenological explorations of multiquark structures~\cite{Yang:2015bmv, Yang:2018oqd, Yang:2020fou, Yang:2020twg, Yang:2021izl}.

Electromagnetic transitions have been treated traditionally within the potential model approach. However, in the last decade, progress has been made using effective field theories (see~\cite{Segovia:2018qzb, Pineda:2013lta} and references therein) and lattice-regularized QCD~\cite{Dudek:2006ej, Becirevic:2012dc}. We shall use the formulae described in Ref.~\cite{Brambilla:2010cs}, but adapting it to our non-relativistic constituent quark model approach. Although such expressions have been used since the early days of hadron spectroscopy, a brief description can be found below. Focusing now on the hadronic transitions, since the energy difference between the initial and final $B_c$ states is expected to be small, the emitted gluons are rather soft. In Ref.~\cite{Gottfried:1977gp}, Gottfried pointed out that this gluon radiation can be treated in a multipole expansion, since the wavelengths of the emitted gluons are large compared to the size of the heavy mesons. The multipole expansion within QCD (QCDME) has been studied by many authors~\cite{Gottfried:1977gp, Bhanot:1979af, Peskin:1979va, Bhanot:1979vb, Voloshin:1978hc, Voloshin:1980zf}, but Yan was the first one to present a gauge-invariant formulation in Refs.~\cite{Yan:1980uh, Kuang:1981se} (see also the interesting advances made very recently in Refs.~\cite{Pineda:2019mhw, TarrusCastella:2021pld}). We shall follow the updated review~\cite{Kuang:2006me} and references therein to calculate the hadronic transitions within our quark model formalism.

The manuscript is arranged as follows. After this introduction, the theoretical framework is presented in Sec.~\ref{sec:Theory}; we explain first the quark model Hamiltonian and then the consistent formulation of radiative and hadronic decays. Section~\ref{sec:Results} is mostly devoted to the analysis and discussion of our theoretical results; we end this section by discussing some strategies for searching for excited $B_c$ mesons and studying their spectroscopy. Finally, we summarize and give some prospects in Sec.~\ref{sec:Epilogue}.

%%%%%%%%%%%%%%%%%%%%%%%%%%%%%%%%%%%%%%%%%%%%%%%%%%%%%%%%%%%%%%%%%%%%%%%%%%%%%%%%
%%%%%%%%%%%%%%%%%%%%%%%%%%%%%%%%%%%%%%%%%%%%%%%%%%%%%%%%%%%%%%%%%%%%%%%%%%%%%%%%

\vspace*{0.25cm}

\section{Theoretical framework}
\label{sec:Theory}

In this section we are going to present, first, a detailed description of all different terms of the interacting potential. Later on, the standard formulae that describes radiative transitions between low-lying $B_c$ states is shown, which includes the dominant E1 and M1 multipole electromagnetic decay rates. And, finally, the latter subsection is dedicated to the hadronic transitions following the QCD multipole expansion method. It consists on a two-step process in which the gluons are first emitted from the heavy quarks and then recombine into light quarks. A multipole expansion of the color gauge field is employed to describe the emission process, whereas the intermediate color octet state is modeled by some sort of quark-antiquark-gluon hybrid wave function.

\subsection{CONSTITUENT QUARK MODEL}
\label{subsec:CQM}

In the heavy quark sector, chiral symmetry is explicitly broken and, thus, the interaction between quarks due to Goldstone-boson exchanges does not take place. Therefore, one-gluon exchange and confinement are the only interactions remaining. The one-gluon exchange potential contains central, tensor and spin-orbit contributions given by
\begin{widetext}
\begin{align}
&
V_{\rm OGE}^{\rm C}(\vec{r}_{ij}) =
\frac{1}{4}\alpha_{s}(\vec{\lambda}_{i}^{c}\cdot
\vec{\lambda}_{j}^{c})\left[ \frac{1}{r_{ij}}-\frac{1}{6m_{i}m_{j}} 
(\vec{\sigma}_{i}\cdot\vec{\sigma}_{j}) 
\frac{e^{-r_{ij}/r_{0}(\mu)}}{r_{ij}r_{0}^{2}(\mu)}\right] \,, \nonumber \\
& 
V_{\rm OGE}^{\rm T}(\vec{r}_{ij})=-\frac{1}{16}\frac{\alpha_{s}}{m_{i}m_{j}}
(\vec{\lambda}_{i}^{c}\cdot\vec{\lambda}_{j}^{c})\left[ 
\frac{1}{r_{ij}^{3}}-\frac{e^{-r_{ij}/r_{g}(\mu)}}{r_{ij}}\left( 
\frac{1}{r_{ij}^{2}}+\frac{1}{3r_{g}^{2}(\mu)}+\frac{1}{r_{ij}r_{g}(\mu)}\right)
\right]S_{ij} \,, \nonumber \\
&
V_{\rm OGE}^{\rm SO}(\vec{r}_{ij})= 
-\frac{1}{16}\frac{\alpha_{s}}{m_{i}^{2}m_{j}^{2}}(\vec{\lambda}_{i}^{c} \cdot
\vec{\lambda}_{j}^{c})\left[\frac{1}{r_{ij}^{3}}-\frac{e^{-r_{ij}/r_{g}(\mu)}}
{r_{ij}^{3}} \left(1+\frac{r_{ij}}{r_{g}(\mu)}\right)\right] \times \nonumber \\ 
& 
\hspace*{1.60cm} \times \left[((m_{i}+m_{j})^{2}+2m_{i}m_{j})(\vec{S}_{+}\cdot\vec{L})+
(m_{j}^{2}-m_{i}^{2}) (\vec{S}_{-}\cdot\vec{L}) \right] \,,
\label{eq:OGEpot}
\end{align}
\end{widetext}
with $\vec{\lambda}^{c}$ being the $SU(3)$ color matrices and $\alpha_{s}$ is
the quark-gluon coupling constant. The regulators $r_{0}(\mu)=\hat{r}_{0}\frac{\mu_{nn}}{\mu_{ij}}$ and $r_{g}(\mu)=\hat{r}_{g}\frac{\mu_{nn}}{\mu_{ij}}$ depend on $\mu_{ij}$ which is the reduced mass of the interacting $q\bar{q}$ pair. The quark tensor operator is $S_{ij}=3(\vec{\sigma}_{i}\cdot\hat{r}_{ij})(\vec{\sigma}_{j}\cdot
\hat{r}_{ij})-\vec{\sigma}_{i}\cdot\vec{\sigma}_{j}$, with $\sigma_i$ denoting the Pauli matrices; and $\vec{S}_{\pm}=\vec{S_{i}}\pm\vec{S}_{j}$ with $S_i=\sigma_i/2$. The contact term of the central potential has been regularized as
\begin{equation}
\delta(\vec{r}_{ij})\sim\frac{1}{4\pi 
r_{0}^{2}}\frac{e^{-r_{ij}/r_{0}}}{r_{ij}}.
\label{eq:delta}
\end{equation}

The wide energy range needed to provide a consistent description of light, strange and heavy mesons requires an effective scale-dependent strong coupling constant. We use the frozen coupling constant~\cite{Vijande:2004he}
\begin{equation}
\alpha_{s}(\mu)=\frac{\alpha_{0}}{\ln\left(
\frac{\mu^{2}+\mu_{0}^{2}}{\Lambda_{0 }^{2}} \right)},
\end{equation}
in which $\mu$ is the reduced mass of the $q\bar{q}$ pair and $\alpha_{0}$, $\mu_{0}$ and $\Lambda_{0}$ are parameters of the model determined by a global fit to the meson spectra.

It is well known that multi-gluon exchanges produce an attractive linearly rising potential proportional to the distance between infinite-heavy quarks. However, sea quarks are also important ingredients of the strong interaction dynamics that contribute to the screening of the rising potential at low momenta and eventually to the breaking of the quark-antiquark binding string~\cite{Bali:2005fu}. Our model tries to mimic this behaviour~\cite{Segovia:2008zza}; the different pieces of the confinement potential are
\begin{align}
&
V_{\rm CON}^{\rm C}(\vec{r}_{ij})=\left[-a_{c}(1-e^{-\mu_{c}r_{ij}})+\Delta
\right] (\vec{\lambda}_{i}^{c}\cdot\vec{\lambda}_{j}^{c}) \,, \nonumber \\
&
V_{\rm CON}^{\rm SO}(\vec{r}_{ij}) =
-(\vec{\lambda}_{i}^{c}\cdot\vec{\lambda}_{j}^{c}) \frac{a_{c}\mu_{c}e^{-\mu_{c}
r_{ij}}}{4m_{i}^{2}m_{j}^{2}r_{ij}} \times \nonumber \\
&
\times 
\left[((m_{i}^{2}+m_{j}^{2})(1-2a_{s}) + 4m_{i}m_{j}(1-a_{s}))(\vec{S}_{+} 
\cdot\vec{L}) \right. \nonumber \\
&
\left. \quad\,\, +(m_{j}^{2}-m_{i}^{2}) (1-2a_{s}) (\vec{S}_{-}\cdot\vec{L})
\right],
\end{align}
where $a_{s}$ controls the mixture between the scalar and vector Lorentz structures of the confinement. At short distances this potential presents a linear behavior with an effective confinement strength $\sigma=-a_{c}\,\mu_{c}\,(\vec{\lambda}^{c}_{i}\cdot \vec{\lambda}^{c}_{j})$, while it becomes constant at large distances. This type of potential shows a threshold defined by
\begin{equation}
V_{\rm thr}=\{-a_{c}+ \Delta\}(\vec{\lambda}^{c}_{i}\cdot
\vec{\lambda}^{c}_{j}).
\end{equation}
No $q\bar{q}$ bound states can be found for energies higher than this threshold. The system suffers a transition from a color string configuration between two static color sources into a pair of static mesons due to the breaking of the color string and the most favored decay into hadrons.

Among the different methods to solve the Schr\"odinger equation in order to  find the quark-antiquark bound states, we use the Gaussian Expansion Method (GEM)~\cite{Hiyama:2003cu} which provides enough accuracy and it simplifies the subsequent evaluation of the needed matrix elements.

This procedure provides the radial wave function solution of the Schr\"odinger equation as an expansion in terms of basis functions
\begin{equation}
R_{\alpha}(r)=\sum_{n=1}^{n_{max}} c_{n}^\alpha \phi^G_{nl}(r),
\end{equation} 
where $\alpha$ refers to the channel quantum numbers. The coefficients, $c_{n}^\alpha$, and the eigenvalue, $E$, are determined from the Rayleigh-Ritz variational principle
\begin{equation}
\sum_{n=1}^{n_{max}} \left[\left(T_{n'n}^\alpha-EN_{n'n}^\alpha\right)
c_{n}^\alpha+\sum_{\alpha'}
\ V_{n'n}^{\alpha\alpha'}c_{n}^{\alpha'}=0\right],
\end{equation}
where $T_{n'n}^\alpha$, $N_{n'n}^\alpha$ and $V_{n'n}^{\alpha\alpha'}$ are the  matrix elements of the kinetic energy, the normalization and the potential,  respectively. $T_{n'n}^\alpha$ and $N_{n'n}^\alpha$ are diagonal, whereas the mixing between different channels is given by $V_{n'n}^{\alpha\alpha'}$.

Following Ref.~\cite{Hiyama:2003cu}, we employ Gaussian trial functions with
ranges  in geometric progression. This facilitates the optimization of ranges
employing a small number of free parameters. Moreover, the geometric
progression is dense at short distances, so that it enables the description of
the dynamics mediated by short range potentials. The fast damping of the
Gaussian tail does not represent an issue, since we can choose the maximal
range much longer than the hadronic size.

Finally, the model parameters can be found in Table~\ref{tab:parameters}. They have been fixed following hadron phenomenology described in, for instance, the literature mentioned previously in the introduction.

\begin{table}[!t]
\caption{\label{tab:parameters} Quark model parameters.}
\begin{ruledtabular}
\begin{tabular}{llc}
Quark masses & $m_{c}$ (MeV) & $1763$ \\
             & $m_{b}$ (MeV) & $5110$ \\
\hline
OGE  & $\alpha_{0}$ & $2.118$ \\
     & $\Lambda_{0}$ $(\mbox{fm}^{-1})$ & $0.113$ \\
     & $\mu_{0}$ (MeV) & $36.976$ \\
     & $\hat{r}_{0}$ (fm) & $0.181$ \\
     & $\hat{r}_{g}$ (fm) & $0.259$ \\
\hline
CON & $a_{c}$ (MeV) & $507.4$ \\
	& $\mu_{c}$ $(\mbox{fm}^{-1})$ & $0.576$ \\
	& $\Delta$ (MeV) & $184.432$ \\
	& $a_{s}$ & $0.81$ \\
\end{tabular}
\end{ruledtabular}
\end{table}

%%%%%%%%%%%%%%%%%%%%%%%%%%%%%%%%%%%%%%%%%%%%%%%%%%%%%%%%%%%%%%%%%%%%%%%%%%%%%%%%

\subsection{Radiative decays}
\label{subsec:radiative}

The decay rate for E1 transitions between an initial state $n^{2S+1}L_J$ and a final state $n'\,^{2S'+1}L'_{J'}$ can be written as
\begin{equation}
\begin{split}
\Gamma_{E1} (n^{2S+1}&L_J  \rightarrow n'\,^{2S'+1}L'_{J'}) = \\
&= \frac{4\alpha e_Q^2 k^3}{3} (2J'+1) S_{fi}^E\,\delta_{SS'}\,|\mathcal
E_{fi}|^2\frac{E_f}{M_i},
\end{split}
\end{equation}
where $e_Q=(e_c m_b - e_b m_c)/(m_c+m_b)$, $k=(M_i^2-M_f^2)/2M_i$ is the emitted photon momentum with $M_i$ ($M_f$) the mass of the initial (final) state, $E_{f}/M_{i}$ is a relativistic correction where $E_f$ the energy of the final state. The statistical factor, $S_{fi}^E$, is given by 
\begin{equation}
S_{fi}^E={\rm max}(L,L')\left\{\begin{matrix} J & 1 & J' \\ L'& S & L
\end{matrix}\right\}^2.
\end{equation}
If the full momentum dependence is retained, the overlap integral, $\mathcal
E_{fi}$, is
\begin{equation}
{\cal E}_{fi} = \frac{3}{k}\int_{0}^{\infty} R_{\alpha'}(r)  
\left[ \frac{kr}{2}j_0\left(\frac{kr}{2}\right)-j_1\left(\frac{kr}{2}\right)
\right] R_{\alpha}(r) \, r^2 \,dr,
\label{eq:EfiE1}
\end{equation}
where $j_{i}(x)$ are the spherical Bessel functions of the first kind and 
$\alpha$ ($\alpha'$) are the initial (final) meson quantum numbers.

The M1 radiative transitions can be evaluated with the following expression
\begin{equation}
\begin{split}
\Gamma_{M1}(n^{2S+1}L_J & \rightarrow n'\,^{2S'+1}L'_{J'}) = \\
&= 
\frac{4\alpha e_Q^2 k^3}{3m_cm_b} (2J'+1)S_{fi}^M |{\mathcal M_{fi}}|^2 \frac{E_f}{M_i},
\end{split}
\end{equation}
where we use the same notation as in the E1 transitions but now 
\begin{equation}
S_{fi}^M=6(2S+1)(2S'+1) \left\{\begin{matrix}J & 1 & J' \\ S'& L & S \end{matrix}\right\}^2 \left\{\begin{matrix}1 & 1/2 & 1/2 \\ 1/2& S' & S \end{matrix}\right\}^2,
\end{equation}
and
\begin{equation}
{\cal M}_{fi} = \int_{0}^{\infty} R_{\alpha'}(r) \, j_{0}\left(\frac{kr}{2}\right) R_{\alpha}(r) \, r^2 \,dr.
\end{equation}

%%%%%%%%%%%%%%%%%%%%%%%%%%%%%%%%%%%%%%%%%%%%%%%%%%%%%%%%%%%%%%%%%%%%%%%%%%%%%%%%

\subsection{Hadronic decays}

One can refer to a hadronic transition in the following general way
\begin{equation}
\Phi_{I} \to \Phi_{F}+h,
\end{equation}
where $h$ denotes the light hadron(s) emerging from the emitted gluons; they are kinematically dominated by either single particle ($\pi^{0}$, $\eta$, $\omega$, $\ldots$) or two particle ($2\pi$, $2K$, $\ldots$) states. The initial and final states of $B_c$ mesons are named as $\Phi_{I}$ and $\Phi_{F}$, respectively.

The emitted gluons are rather soft because the energy difference between the initial and final charm-beauty states is small. Gottfried pointed out in Ref.~\cite{Gottfried:1977gp} that the gluon radiation can be expanded in multipoles since the wavelengths of emitted gluons are larger than the size of $B_c$-meson states. After the expansion of the gluon field, the Hamiltonian of the system can be decomposed as
\begin{equation}
{\cal H}^{\rm eff}_{\rm QCD} = {\cal H}^{(0)}_{\rm QCD} + {\cal H}^{(1)}_{\rm
QCD} + {\cal H}^{(2)}_{\rm QCD},
\label{eq:Hqcd}
\end{equation}
with ${\cal H}^{(0)}_{\rm QCD}$ the sum of the kinetic and potential energies of the bottom-charmed meson, and ${\cal H}^{(1)}_{\rm QCD}$ and ${\cal H}^{(2)}_{\rm
QCD}$ are defined by 
\begin{equation}
\begin{split}
{\cal H}^{(1)}_{\rm QCD} &= Q_{a} A^{a}_{0}(x,t), \\
{\cal H}^{(2)}_{\rm QCD} &=-d_{a} E^{a}(x,t) - m_{a} B^{a}(x,t),
\label{eq:Hqcd2}
\end{split}
\end{equation}
in which $Q_{a}$ is the color charge and the color electric and magnetic dipole moments are represented by $d_{a}$ and $m_{a}$, respectively. Since we are working with $c\bar{b}$-pairs that form a color singlet object, there is no contribution from the ${\cal H}^{(1)}_{\rm QCD}$ and only $E_{l}$ and $B_{m}$ transitions can take place.

A multipole expansion within QCD is now necessary in order to continue with the computation of the hadronic transitions between $B_c$ staes. A brief description of the derived formulae following the updated review~\cite{Kuang:2006me} can be found below.

\subsubsection{Spin-nonflip $\pi\pi$ and $\eta$ transitions}
\label{subsubsec:Spinnonflip}

The spin-nonflip $\pi\pi$-decay is dominated by the double electric-dipole term (E1-E1) in the QCD multipole expansion, and thus the transition amplitude can be written as follows
\begin{equation}
{\cal M}_{E1E1}=i\frac{g_{E}^{2}}{6} \left\langle\right.\!\! \Phi_{F}h \,
|\vec{x}\cdot\vec{E} \, \frac{1}{E_{I}-H^{(0)}_{QCD}-iD_{0}} \,
\vec{x}\cdot\vec{E}| \, \Phi_{I} \!\! \left.\right\rangle,
\label{eq:E1E1}
\end{equation}
where $\vec{x}$ is the separation between the $c$-quark and $\bar{b}$-antiquark, and $(D_0)_{bc}\equiv\delta_{bc}\partial_{0}-g_{s}f_{abc}A^{a}_{0}$.

Inserting a complete set of intermediate states, the transition amplitude, Eq.~\eqref{eq:E1E1}, becomes
\begin{equation}
{\cal M}_{E1E1}=i\frac{g_{E}^{2}}{6} \sum_{KL}
\frac{\left\langle\right.\!\! \Phi_{F}|x_k|KL \!\!\left.\right\rangle
\left\langle\right.\!\! KL|x_l|\Phi_I \!\!\left.\right\rangle}{E_I-E_{KL}}
\left\langle\right.\!\! \pi\pi|E^{a}_{k} E^{a}_{l}|0 \!\!\left.\right\rangle,
\label{eq:factorizedE1E1}
\end{equation}
where $E_{KL}$ is the energy eigenvalue of the intermediate state $|KL\rangle$ with the principal quantum number $K$ and the orbital angular momentum $L$.

The intermediate states in the hadronic transition can be considered as hybrid mesons consisting on a color-octet $c\bar{b}$-pair plus gluon(s). They are very hard to calculate in QCD from first principles when the quark-antiquark pair is open-flavor; however, it is worth mentioning herein that there exist non-relativistic effective field theories~\cite{Berwein:2015vca, Brambilla:2018pyn, Brambilla:2019jfi} and lattice-regularised QCD~\cite{HadronSpectrum:2012gic, Cheung:2016bym, Ryan:2020iog} computations of, at least, the first multiplet of quark-gluon hybrid mesons when the quark and antiquark are of the same heavy flavor. We shall take a reasonable model which has been already used for the study of similar hadronic transitions in the charmonium and bottomonium sectors~\cite{Segovia:2014mca, Segovia:2015raa, Segovia:2016xqb}, and it will be explained below.

One can see in Eq.~(\ref{eq:factorizedE1E1}) that the transition amplitude splits into two factors. The first one concerns to the wave functions and energies of the initial and final quarkonium states as well as those of the intermediate hybrid mesons. All these quantities can be calculated using suitable quark-gluon models. The second one describes the conversion of the emitted gluons into light hadrons. As the momenta involved are very low, this matrix element cannot be calculated using perturbative QCD and one needs to resort to a phenomenological approach based on soft-pion techniques~\cite{Brown:1975dz}. In the center-of-mass frame, the two pion momenta $q_{1}$ and $q_{2}$ are the only independent variables describing this matrix element which, in the nonrelativistic limit, can be parametrized as~\cite{Brown:1975dz, Yan:1980uh, Kuang:1981se, Kuang:2006me}
\begin{equation}
\begin{split}
& \frac{g_{E}^{2}}{6} \left\langle\right.\!\!
\pi_{\alpha}(q_{1})\pi_{\beta}(q_{2})|E^{a}_{k}E^{a}_{l}|0
\!\!\left.\right\rangle =
\frac{\delta_{\alpha\beta}}{\sqrt{(2\omega_{1})(2\omega_ {2})}} \,\times \\
&
\times
\left[C_{1}\delta_{kl}q^{\mu}_{1}q_{2\mu} + C_{2}\left(q_{1k}q_{2l}+q_{1l}q_{2k}
-\frac{2}{3}\delta_{kl}\vec{q}_{1}\cdot\vec{q}_{2}\right)\right],
\label{HofE1E1}
\end{split}
\end{equation}
where $C_{1}$ and $C_{2}$ are two unknown constants, related to our ignorance about the mechanism of the conversion of the emitted gluons into light hadron(s). The $C_{1}$ term is isotropic, while the $C_{2}$ term has a $L=2$ angular dependence. Thus, $C_{1}$ is involved in hadronic transitions where $\Delta l = l_f-l_i = 0$, while $C_{2}$ begins to participate when $\Delta l = 2$. 

It is also important to mention here that the above parameters are considered theoretically as Wilson coefficients and thus they depend on the characteristic energy scale of the physical process. They have been fixed in our previous studies of hadronic transitions within the charmonium and bottomonium sectors~\cite{Segovia:2014mca, Segovia:2015raa} and, in order to gain predictive power, we use here the values corresponding to the bottomonium case.

Finally, the transition rate is given by
\begin{widetext}
\begin{equation}
\begin{split}
\Gamma\left(\Phi_{I}(^{2s+1}{l_{I}}_{J_{I}}) \to
\Phi_{F}(^{2s+1}{l_{F}}_{J_{F}}) + \pi\pi\right) = &
\delta_{l_{I}l_{F}}\delta_{J_{I}J_{F}} (G|C_{1}|^{2}-\frac{2}{3}H|C_{2}|^{2}
)\left|\sum_{L}(2L+1) \left(\begin{matrix} l_{I} & 1 & L \\ 0 & 0 & 0
\end{matrix}\right) \left(\begin{matrix} L & 1 & l_{I} \\ 0 & 0 & 0
\end{matrix}\right) f_{IF}^{L11}\right|^{2} \\
&
+(2l_{I}+1)(2l_{F}+1)(2J_{F}+1) \sum_{k} (2k+1) (1+(-1)^{k})
\left\lbrace\begin{matrix} s & l_{F} & J_{F} \\ k & J_{I} & l_{I}
\end{matrix}\right\rbrace^{2} H |C_{2}|^{2} \times \\
&
\times\left|\sum_{L} (2L+1) \left(\begin{matrix} l_{F} & 1 & L \\ 0 & 0 & 0
\end{matrix}\right) \left(\begin{matrix} L & 1 & l_{I} \\ 0 & 0 & 0
\end{matrix}\right) \left\lbrace\begin{matrix} l_{I} & L & 1 \\ 1 & k & l_{F}
\end{matrix}\right\rbrace f_{IF}^{L11} \right|^{2},
\label{eq:gamapipi}
\end{split}
\end{equation}
with
\begin{equation}
f_{IF}^{LP_{I}P_{F}} = \sum_{K} \frac{1}{M_{I}-M_{KL}} \left[\int dr\,
r^{2+P_{F}} R_{F}(r)R_{KL}(r)\right] \left[\int dr' r'^{2+P_{I}} R_{KL}(r')
R_{I}(r')\right].
\label{eq:fifl}
\end{equation}
\end{widetext}
$R_{KL}(r)$ is the radial wave function of the intermediate quark-gluon states, whereas $R_{I}(r)$ and $R_{F}(r)$ are the radial wave functions of the initial and final states, respectively. The mass of the decaying meson is $M_{I}$, whereas the ones corresponding to the hybrid states are $M_{KL}$. The quantities $G$ and $H$ are phase-space integrals
\begin{equation}
\begin{split}
G=&\frac{3}{4}\frac{M_{F}}{M_{I}}\pi^{3}\int
dM_{\pi\pi}^{2}\,k\,\left(1-\frac{4m_{\pi}^{2}}{M_{\pi\pi}^{2}}\right)^{1/2}(M_{
\pi\pi}^{2}-2m_{\pi}^{2})^{2}, \\
H=&\frac{1}{20}\frac{M_{F}}{M_{I}}\pi^{3}\int
dM_{\pi\pi}^{2}\,k\,\left(1-\frac{4m_{\pi}^{2}}{M_{\pi\pi}^{2}}\right)^{1/2}
\times \\
&
\times\left[(M_{\pi\pi}^{2}-4m_{\pi}^{2})^{2}\left(1+\frac{2}{3}\frac{k^{2}}{M_{
\pi\pi}^{2}}\right)\right. \\
&
\left.\quad\,\, +\frac{8k^{4}}{15M_{\pi\pi}^{4}}(M_{\pi\pi}^{4}+2m_{\pi}^{2}
M_{\pi\pi}^{2}+6m_{\pi}^{4})\right],
\end{split}
\end{equation}
with the momentum $k$ given by
\begin{equation}
k = \frac{\sqrt{\left[(M_{I}+M_{F})^{2}-M_{\pi\pi}^{2}\right]
\left[(M_{I}-M_{F})^{2}-M_{ \pi\pi}^{2}\right]}}{2M_{I}}.
\end{equation}

The leading multipoles of spin-nonflip $\eta$-transitions between spin-triplet
$S$-wave states are M1-M1 and E1-M2. Therefore, the matrix element is given
schematically by
\begin{equation}
{\cal M}(^{3}S_{1}\to\,^{3}\!S_{1} + \eta) = {\cal M}_{M1M1} +
{\cal M}_{E1M2}.
\end{equation}
After some algebra and assuming that ${\cal M}_{M1M1}=0$ (see
Ref.~\cite{Kuang:1981se} for details), the decay rate can be written as
\begin{equation}
\Gamma(\Phi_{I}(^{3}S_{1}) \to \Phi_{F}(^{3}S_{1}) + \eta) =
\frac{8\pi^{2}}{27} \frac{M_{f}C_{3}^{2}}{M_{i}m_{Q}m_{Q'}} |f_{IF}^{111}|^{2}
|\vec{q}\,|^{3},
\end{equation}
where $\vec{q}$ is the momentum of $\eta$, the function $f_{IF}^{111}$ is defined in Eq.~(\ref{eq:fifl}), and $C_{3}$ is a new parameter.
%which should be fixed through the $\Upsilon(2S)\to\Upsilon(1S)\eta$ reaction.

\subsubsection{Spin-flip $\pi\pi$ and $\eta$ transitions}
\label{subsubsec:spinflip}

The spin-flip $\pi\pi$- and $\eta$-transitions between $B_c$-mesons are induced by an E1-M1 multipole amplitude. Within the hadronization approach presented above, the description of this kind of decays implies the introduction of another phenomenological constant which should be fixed by experiment. Therefore, as one can deduce, the decay model for hadronic transitions begins to loose its predictive power.

In order to avoid this undesirable feature, the term which describes the conversion of the emitted gluons into light hadrons can be computed assuming a duality argument between the physical light hadron final state and the associated two-gluon final state~\cite{Kuang:1981se}:
\begin{equation}
\begin{split}
\Gamma(\Phi_{I}\to \Phi_{F} + \pi\pi) &\sim \Gamma(\Phi_{I}\to \Phi_{F}gg), \\
\Gamma(\Phi_{I}\to \Phi_{F} + \eta) &\sim \Gamma(\Phi_{I}\to 
\Phi_{F}(gg)_{0^{-}}), \\
\end{split}
\end{equation}
where in the second line the two gluons are projected into a $J^{P}=0^{-}$ state to simulate the $\eta$-meson. The advantage of this approach is that we have now only two free parameters, $g_{E}$ and $g_{M}$, in order to fix the spin-nonflip and spin-flip $\pi\pi$- and $\eta$-hadronic transitions. The values used herein are those reported in Ref.~\cite{Segovia:2015raa}.

Explicit expressions within this new approach of the decay rates for the spin-nonflip $\pi\pi$- and $\eta$-transitions can be found in Refs.~\cite{Kuang:1981se, Kuang:2006me}. The decay rates for the spin-flip $\pi\pi$- and $\eta$-transitions are
\begin{equation}
\begin{split}
&
\Gamma(\Phi_{I}(^{3}{l_{I}}_{J_{I}})\to \Phi_{F}(^{1}{l_{F}}_{J_{F}}) +
\pi\pi) = \frac{g_{E}^2g_{M}^2}{36m_{Q}m_{Q'}} \times \\
&
\hspace*{0.40cm} \times \frac{(M_{I}-M_{F})^{7}}{315\pi^{3}} (2l_{F}+1)
\left(\begin{matrix} l_{F} & 1 & l_{I} \\ 0 & 0 & 0 \end{matrix}\right)^{2}
|f_{IF}^{l_{F}10}+f_{IF}^{l_{I}01}|^{2}, \\[2ex]
&
\Gamma(\Phi_{I}(^{3}{S}_{J_{I}})\to \Phi_{F}(^{1}{P}_{J_{F}}) + \eta) =
\frac{g_{M}^{2}}{g_{E}^{2}} \frac{E_{F}}{M_{I}} |\vec{q}\,| \times \\
&
\hspace*{0.40cm} \times \frac{\pi}{1144m_{Q}m_{Q'}} 
\left(\frac{4\pi}{\sqrt{6}}f_{\pi}m_{\eta}^{2}\right)^{2}
|f_{IF}^{110}+f_{IF}^{001}|^{2}.
\label{eq:spinflip}
\end{split}
\end{equation}
The decay rate of the spin-flip $\eta$-transition in Eq.~(\ref{eq:spinflip}) can be read from the decay rate of the the isospin violating hadronic transition~\cite{Kuang:2006me}
\begin{equation}
\begin{split}
&
\Gamma(\Phi_{I}(^{3}{S}_{J_{I}})\to \Phi_{F}(^{1}{P}_{J_{F}}) + \pi^{0}) =
\frac{g_{M}^{2}}{g_{E}^{2}} \frac{E_{F}}{M_{I}} |\vec{q}\,| \times \\
&
\hspace*{0.40cm} \times \frac{\pi}{1144m_{Q}m_{Q'}} 
\left(\frac{4\pi}{\sqrt{2}}\frac{m_{d}-m_{u}}{m_{d}+m_{u}}f_{\pi}m_{\pi}^{2}
\right)^{2} |f_{IF}^{110}+f_{IF}^{001}|^{2},
\end{split}
\end{equation}
in which the factor $(m_{d}-m_{u})/(m_{d}+m_{u})\approx0.35$ reflects the violation of isospin.

%%%%%%%%%%%%%%%%%%%%%%%%%%%%%%%%%%%%%%%%%%%%%%%%%%%%%%%%%%%%%%%%%%%%%%%%%%%%%%%%

\subsubsection{A model for hybrid mesons}
\label{subsubsec:hybrids}

One might expect to have bound states in which the gluon field itself is excited and carries $J^{PC}$ quantum numbers. Quantum Chromodynamics does not forbid this and, in fact, it should expected from its general properties. The gluonic quantum numbers couple to those of the quark-antiquark pair, giving rise to the so-called exotic $J^{PC}$ mesons, but also can produce hybrid mesons with natural quantum numbers. We are interested on the last ones because they are involved in the calculation of hadronic transitions within the QCDME approach.

An extension of the non-relativistic constituent quark model described above to include hybrid states was presented in~\cite{Segovia:2014mca}  (see also Refs.\cite{Segovia:2015raa, Segovia:2016xqb}). This extension is inspired on the Buchmuller-Tye quark-confining string (QCS) model~\cite{Tye:1975fz,  Giles:1977mp, Buchmuller:1979gy} in which the meson is composed of a quark and antiquark linked by an appropriate color electric flux line (the string).

The string can carry energy-momentum only in the region between the quark and
the antiquark. The string and the quark-antiquark pair can rotate as a unit and
also vibrate. Ignoring its vibrational motion, the equation which describes the
dynamics of the quark-antiquark pair linked by the string should be the usual
Schr\"odinger equation with a confinement potential. Gluon excitation effects
are described by the vibration of the string. These vibrational modes provide
new states beyond the naive meson picture.

The coupled equations that describe the dynamics of the string and the quark
sectors are very non-linear so that there is no hope of solving them completely. Using the Bohr-Sommerfeld quantization, the vibrational potential energy can be estimated as a function of the interquark distance and then, via the Bohr-Oppenheimer method, these vibrational energies are inserted into the meson equation as an effective potential, $V_{n}(r)$.

\begin{table}[!t]
\begin{center}
\begin{tabular}{cccc}
\hline
\hline
\tstrut
K & $L=0$ & $L=1$ & $L=2$ \\
\hline
\tstrut
$1$  & $7328$ & $7567$ & $7733$ \\
$2$  & $7667$ & $7828$ & $7956$ \\
$3$  & $7910$ & $8034$ & $8136$ \\
$4$  & $8102$ & $8199$ & $8281$ \\
$5$  & $8255$ & $8333$ & $8399$ \\
$6$  & $8378$ & $8441$ & $8493$ \\
$7$  & $8477$ & $8525$ & $8566$ \\
$8$  & $8553$ & $8588$ & - \\
\hline
\multicolumn{4}{c}{Threshold = 8595 MeV} \\
\hline
\hline
\end{tabular}
\caption{\label{tab:hybridsbb} Hybrid meson masses, in MeV, calculated in the
$c\bar{b}$ sector. The variation of the parameter $\alpha_{n}$ which range
between $1<\alpha_{n}<\sqrt{2}$ modifies the energy as much as $30\,{\rm MeV}$,
we have taken $\alpha_{n}=\sqrt{1.5}$.}
\end{center}
\end{table}

\begin{table*}[!t]
\begin{ruledtabular}
\caption{\label{tab:predmassesbc} Predicted masses, in MeV, of the $B_c$ states which are expected to be either below or around $BD$-threshold. All spin and orbital partial waves compatible with total spin and parity quantum numbers are considered in the coupled-channels Schr\"odinger equation and, thus, the fourth column indicates the dominant channel. We compare with available experimental data~\cite{ParticleDataGroup:2020ssz}, recent lattice QCD studies~\cite{Mathur:2018epb, Dowdall:2012ab} and some other model predictions~\cite{Godfrey:1985xj, Ebert:2002pp, Eichten:1994gt}.}
\begin{tabular}{ccccccccccc}
State & $J^{P}$ & $n$ & $^{2S+1}L_J$ & The. & Exp.~\cite{ParticleDataGroup:2020ssz} & Ref.~\cite{Mathur:2018epb} & Ref.~\cite{Dowdall:2012ab} & Ref.~\cite{Godfrey:1985xj} & Ref.~\cite{Ebert:2002pp} & Ref.~\cite{Eichten:1994gt} \\
\hline
$B_c$ & $0^-$ & $1$ & $^{1}S_0$ & $6277$ & $6274.47\pm0.32$ & $6276\pm7$ & $6278\pm9$ & $6271$ & $6270$ & $6264$ \\
& & $2$ & $^{1}S_0$ & $6868$ & $6871.2\pm1.0$ & - & $6894\pm21$ & $6855$ & $6835$ & $6856$ \\ 
\hline
$B_{c0}^\ast$ & $0^+$ & $1$ & $^{3}P_0$ & $6689$ & - & $6712\pm19$ & $6707\pm16$ & $6706$ & $6699$ & $6700$ \\
& & $2$ & $^{3}P_0$ & $7109$ & - & - & - & $7122$ & $7091$ & $7108$ \\
\hline
$B_{c}^\ast$ & $1^-$ & $1$ & $^{3}S_1$ & $6328$ & - & $6331\pm7$ & $6332\pm9$ & $6338$ & $6332$ & $6337$ \\
& & $2$ & $^{3}S_1$ & $6898$ & - & - & $6922\pm21$ & $6887$ & $6881$ & $6899$ \\
& & $3$ & $^{3}D_1$ & $6999$ & - & - & - & $7028$ & $7072$ & $7012$ \\
\hline
$B_{c1}$ & $1^+$ & $1$ & $^{3}P_1$ & $6723$ & - & $6736\pm18$ & $6742\pm16$ & $6741$ & $6734$ & $6730$ \\
& & $2$ & $^{1}P_1$ & $6731$ & - & - & - & $6750$ & $6749$ & $6736$ \\
& & $3$ & $^{3}P_1$ & $7135$ & - & - & - & $7145$ & $7126$ & $7135$ \\
& & $4$ & $^{1}P_1$ & $7142$ & - & - & - & $7150$ & $7145$ & $7142$ \\
\hline
$B_{c2}$ & $2^-$ & $1$ & $^{1}D_2$ & $7002$ & - & - & - & $7036$ & $7079$ & $7009$ \\
& & $2$ & $^{3}D_2$ & $7011$ & - & - & - & $7041$ & $7077$ & $7012$ \\
\hline
$B_{c2}^\ast$ & $2^+$ & $1$ & $^{3}P_2$ & $6742$ & - & - & - & $6768$ & $6762$ & $6747$ \\
&       & $2$ & $^{3}P_2$ & $7151$ & - & - & - & $7164$ & $7156$ & $7153$ \\
\hline
$B_{c3}^\ast$ & $3^-$ & $1$ & $^{3}D_3$ & $7009$ & - & - & - & $7045$ & $7081$ & $7005$ \\
\hline
\hline
\multicolumn{10}{c}{BD-threshold = $7144-7149\,\text{MeV}$~\cite{ParticleDataGroup:2020ssz}}
\end{tabular}
\end{ruledtabular}
\end{table*}

Therefore, the potential for hybrid mesons derived from our non-relativistic constituent quark model has the following expression:
\begin{equation}
V_{\rm hyb}(r)=V_{\rm OGE}^{\rm C}(r) + V_{\rm CON}^{\rm C}(r) +
\left[V_{n}(r) - \sigma(r)r\right],
\label{eq:pothyb}
\end{equation}
where $V_{\rm OGE}^{\rm C}(r) + V_{\rm CON}^{\rm C}(r)$ would be the naive quark-antiquark potential, $V_{n}(r)$ the vibrational one, and the definition of $\sigma(r)$ is
\begin{equation}
\sigma(r) = \frac{16}{3} a_c \left( \frac{1-e^{-\mu_c r}}{r} \right) \,.
\end{equation}
We must subtract the term $\sigma(r)r$ because it appears twice, one in $V_{\rm CON}^{\rm C}(r)$ and the other one in $V_{n}(r)$. This potential does not include new model parameters and depends  only on those coming from the original quark model. In this sense, the calculation of the hybrid states is parameter-free. More explicitly, our different contributions are
\begin{equation}
\begin{split}
V_{\rm OGE}^{\rm C}(r) &= -\frac{4\alpha_{s}}{3r}, \\
V_{\rm CON}^{\rm C}(r) &= \frac{16}{3} [a_{c}(1-e^{-\mu_{c}r})-\Delta], \\
V_{n}(r) &= \sigma(r)r \left\lbrace 1 + \frac{2n\pi}{\sigma(r)
\left[(r-2d)^{2}+4d^{2}\right]} \right\rbrace^{1/2},
\end{split}
\end{equation}
where the vibrational potential energy can be estimated using the Bohr-Sommerfeld quantization and assuming the quark mass to be very heavy so that the ends of the string are fixed~\cite{Giles:1977mp}. In order to relax the last assumption one can define a parameter $d$ given by
\begin{equation}
d(m_{Q},r,\sigma,n) = \frac{\sigma(r)r^{2}\alpha_{n}}{4(m_{Q}+m_{Q'}+\sigma(r)r\alpha_{n})}.
\end{equation}
in which $\alpha_{n}$ relates to the shape of the vibrating string~\cite{Giles:1977mp}, and can take the values $1\leq\alpha_{n}^{2}\leq2$.

An important feature of our hybrid model is that, just like the naive quark model, the hybrid potential has a threshold from which no more states can be found and so we have a finite number of hybrid states in the spectrum. Hybrid meson masses calculated in the $B_c$ sector using our model are shown in Table~\ref{tab:hybridsbb}.

%%%%%%%%%%%%%%%%%%%%%%%%%%%%%%%%%%%%%%%%%%%%%%%%%%%%%%%%%%%%%%%%%%%%%%%%%%%%%%%%
%%%%%%%%%%%%%%%%%%%%%%%%%%%%%%%%%%%%%%%%%%%%%%%%%%%%%%%%%%%%%%%%%%%%%%%%%%%%%%%%

\section{Results}
\label{sec:Results}

Table~\ref{tab:predmassesbc} shows the predicted masses of the low-lying $B_c$ states which are expected to be either below or around $BD$-threshold ($7144-7149\,\text{MeV}$)~\cite{ParticleDataGroup:2020ssz}. One can see that there are two $S$-wave multiplets with spin-parity $0^-$ and $1^-$; another two $P$-wave multiplets with quantum numbers $J^P=0^+$, $1^+$ and $2^+$; one $D$-wave multiplet with $J^P=1^-$, $2^-$ and $3^-$; and one $F$-wave multiplet with $J^P=2^+$ very close to $BD$-threshold. The proliferation of states in the spin-parity channels $1^+$ and $2^-$ is due to the coupling of the $S=0$ and $S=1$ channels given by the anti-symmetric spin-orbit term of the quark--anti-quark potential.

We compare our results with the scarce experimental data collected by the PDG~\cite{ParticleDataGroup:2020ssz}. These experimental results only cover the lowest-lying states of the $J^P=0^-$ sector. To compare other sectors we included recent lattice QCD studies, such as the quenched $2+1$~\cite{Allison:2004be} and the $2+1+1$ flavors~\cite{Dowdall:2012ab} calculations of the HPQCD Collaboration, and the $2+1+1$ flavors analysis of Ref.~\cite{Mathur:2018epb}. An overall good agreement with the available lattice/experimental data for the $B_c$ spectra below the lowest $BD$ threshold is obtained. Finally, our predicted masses are also compared with those obtained by a significant sample of phenomenological models~\cite{Godfrey:1985xj, Ebert:2002pp, Eichten:1994gt}. Within the expected theoretical accuracy, the different models are in remarkable agreement for most part of the spectrum. The spin-dependent splittings are also in reasonable agreement; the only significant difference is the larger spread $(\approx 70\,\text{MeV})$ for the $1D$ multiplet centre of gravity predictions. Potential models can therefore be used as a reliable guide in searching for the $B_c$ excited states.

Above the aforementioned $BD$-threshold, coupled-channels effects may appear. The influence of coupling bare $c\bar b$ states with open channels depends on the relative position of the $c\bar b$ mass and the open threshold. When the value of the threshold energy $E$ is greater than the $q\bar q$ mass $M$, the effective potential is repulsive and it is unlikely that the coupling can generate a bound state rather than a dressing effect of the bare state. However, if $M>E$ the potential becomes negative, an extra bound state with a large molecular probability may appear. All this is explained in Ref.~\cite{Ortega:2020uvc} where an example of the influence of two-meson thresholds on the $B_c$ states in the $J^P=0^+$, $1^+$ and $2^+$ channels is shown.

%%%%%%%%%%%%%%%%%%%%%%%%%%%%%%%%%%%%%%%%%%%%%%%%%%%%%%%%%%%%%%%%%%%%%%%%%%%%%%%%

\begin{table}[!t]
\begin{ruledtabular}
\caption{\label{tab:E1-1} The radiative E1 electromagnetic transitions for dominant $S$-wave states. We compare with some other model predictions~\cite{Godfrey:2004ya, Ebert:2002pp, Eichten:1994gt}.}
\begin{tabular}{ccrrrr}
Initial & Final & $\Gamma_{\text{The.}}$ & Ref.~\cite{Godfrey:2004ya} & Ref.~\cite{Ebert:2002pp} & Ref.~\cite{Eichten:1994gt} \\
state & state & (keV) & & & \\
\hline
$B_c^\ast(2S)$ & $\gamma B_{c0}^\ast(1P)$ & $8.8$ & $2.9$ & $3.78$ & $7.8$ \\
               & $\gamma B_{c1}(1P)$ & $20$ & $4.7$ & $5.05$ & $14.5$ \\
               & $\gamma B_{c1}(2P)$ & $1.5\times10^{-3}$ & $0.7$ & $0.63$ & $0.0$ \\
               & $\gamma B_{c2}^\ast(1P)$ & $29$ & $5.7$ & $5.18$ & $17.7$ \\[2ex]
$B_c(2S)$ & $\gamma B_{c1}(1P)$ & $4.8\times10^{-3}$ & $1.3$ & $1.02$ & $0.0$ \\
          & $\gamma B_{c1}(2P)$ & $35$ & $6.1$ & $3.72$ & $5.2$ \\
\end{tabular}
\end{ruledtabular}
\end{table}

Our predictions for the radiative E1 electromagnetic transitions for dominant $S$-wave states are shown in Table~\ref{tab:E1-1}. Since the $B_c^{(\ast)}(2S)$ states have been already seen by the ATLAS~\cite{Sirunyan:2019osb} and LHCb~\cite{Aaij:2019ldo} experiments at CERN, they can be the gateway for the exploration of first and second $P$-wave multiplets. In fact, the $B_c^\ast(2S)$ state has partial widths ranging from a few keV to tens of keV, and the $B_c(2S)$ has a decay rate of $35\,\text{keV}$. We compare our results with those from some other model predictions~\cite{Godfrey:2004ya, Ebert:2002pp, Eichten:1994gt}; in general, ours are larger than those of Refs.~\cite{Godfrey:2004ya, Ebert:2002pp} and of the same order of magnitude than the ones collected in Ref.~\cite{Eichten:1994gt}. The differences are associated with both quark model assumptions and solving, or not, a coupled-channels Schr\"odinger equation, being the work reported in Ref.~\cite{Eichten:1994gt} closer to ours. As one can see, the decay rates are sensible to the mixing between different partial waves in a given wave function and such mixing is completely fixed in our computation through the tensor and the antisymmetric spin-orbit potentials, which are solved non-perturbatively through their exact treatment in the Schr\"odinger equation.

\begin{table}[!t]
\begin{ruledtabular}
\caption{\label{tab:E1-2} The radiative E1 electromagnetic transitions for dominant $P$-wave states. We compare with some other model predictions~\cite{Godfrey:2004ya, Ebert:2002pp, Eichten:1994gt}.}
\begin{tabular}{ccrrrr}
Initial & Final & $\Gamma_{\text{The.}}$ & Ref.~\cite{Godfrey:2004ya} & Ref.~\cite{Ebert:2002pp} & Ref.~\cite{Eichten:1994gt} \\
state & state & (keV) & & & \\
\hline
$B_{c0}^\ast(1P)$ & $\gamma B_{c}^\ast(1S)$ & $119$ & $55$ & $67.2$ & $79.2$ \\[2ex]
$B_{c0}^\ast(2P)$ & $\gamma B_{c}^\ast(1S)$ & $28$ & $1$ & - & $21.9$ \\
                  & $\gamma B_{c}^\ast(2S)$ & $77$ & $42$ & $29.2$ & $41.2$ \\
                  & $\gamma B_{c}^\ast(1D)$ & $17$ & $4.2$ & $0.036$ & $6.9$ \\[2ex]
$B_{c1}(1P)$ & $\gamma B_{c}(1S)$ & $1.5\times10^{-4}$ & $13$ & $18.4$ & $0.0$ \\
             & $\gamma B_{c}^\ast(1S)$ & $146$ & $60$ & $78.9$ & $99.5$ \\[2ex]
$B_{c1}(2P)$ & $\gamma B_{c}(1S)$ & $173$ & $80$ & $132$ & $56.4$ \\
             & $\gamma B_{c}^\ast(1S)$ & $1.8\times10^{-3}$ & $11$ & $13.6$ & $0.1$ \\[2ex]
$B_{c1}(3P)$ & $\gamma B_{c}(1S)$ & $1.4$ & - & - & - \\
             & $\gamma B_{c}(2S)$ & $2.4$ & - & - & - \\
             & $\gamma B_{c}^\ast(1S)$ & $50$ & - & - & - \\
             & $\gamma B_{c}^\ast(2S)$ & $88$ & - & - & - \\
             & $\gamma B_{c}^\ast(1D)$ & $6.4$ & - & - & - \\
             & $\gamma B_{c2}(1D)$ & $14$ & - & - & - \\
             & $\gamma B_{c2}(2D)$ & $5.1$ & - & - & - \\[2ex]
%             & $\gamma B_{c3}^\ast(1D)$ & & & & \\
%
$B_{c1}(4P)$ & $\gamma B_{c}(1S)$ & $79$ & - & - & - \\
             & $\gamma B_{c}(2S)$ & $101$ & - & - & - \\
             & $\gamma B_{c}^\ast(1S)$ & $1.5$ & - & - & - \\
             & $\gamma B_{c}^\ast(2S)$ & $1.9$ & - & - & - \\
             & $\gamma B_{c}^\ast(1D)$ & $0.14$ & - & - & - \\
             & $\gamma B_{c2}(1D)$ & $10$ & - & - & - \\
             & $\gamma B_{c2}(2D)$ & $15$ & - & - & - \\[2ex]
%             & $\gamma B_{c3}^\ast(1D)$ & & & & \\
%
$B_{c2}^\ast(1P)$ & $\gamma B_{c}^\ast(1S)$ & $156$ & $83$ & $107$ & $112.6$ \\[2ex]
$B_{c2}^\ast(2P)$ & $\gamma B_{c}^\ast(1S)$ & $67$ & $14$ & - & $25.8$ \\
                  & $\gamma B_{c}^\ast(2S)$ & $96$ & $55$ & $57.3$ & $73.8$ \\
                  & $\gamma B_{c}^\ast(1D)$ & $0.27$ & $0.1$ & $0.035$ & $0.2$ \\
                  & $\gamma B_{c2}(1D)$ & $2.4$ & $0.7$ & $0.113$ & - \\
                  & $\gamma B_{c2}(2D)$ & $1.9$ & $0.6$ & $0.269$ & $3.2$ \\
                  & $\gamma B_{c3}^\ast(1D)$ & $24$ & $6.8$ & $1.59$ & $17.8$ \\
\end{tabular}
\end{ruledtabular}
\end{table}

Table~\ref{tab:E1-2} compare our results on the radiative E1 electromagnetic transitions for dominant $P$-wave states with those of Refs.~\cite{Godfrey:2004ya, Ebert:2002pp, Eichten:1994gt}. One can observe that the differences between models are less cumbersome, although they still exist; and, again, our results are in better agreement with those of Ref.~\cite{Eichten:1994gt}. Table~\ref{tab:E1-2} also shows that there are radiative E1 electromagnetic transitions from $P$-wave to $S$-wave states that have rates of the order of tens to hundreds keV. Some remarkable examples are the reactions in which the $P$-wave states decay to $B_c(nS)$ and $B_c^\ast(nS)$ mesons, and thus making these transitions the most feasible ones to be explored by experiments in the near future.

\begin{table}[!t]
\begin{ruledtabular}
\caption{\label{tab:E1-3} The radiative E1 electromagnetic transitions for dominant $D$-wave states. We compare with some other model predictions~\cite{Godfrey:2004ya, Ebert:2002pp, Eichten:1994gt}.}
\begin{tabular}{ccrrrr}
Initial & Final & $\Gamma_{\text{The.}}$ & Ref.~\cite{Godfrey:2004ya} & Ref.~\cite{Ebert:2002pp} & Ref.~\cite{Eichten:1994gt} \\
state & state & (keV) & & & \\
\hline
$B_c^\ast(1D)$ & $\gamma B_{c0}^\ast(1P)$ & $98$ & $55$ & $128$ & $88.6$ \\
               & $\gamma B_{c1}(1P)$ & $64$ & $28$ & $73.8$ & $49.3$ \\
               & $\gamma B_{c1}(2P)$ & $1.3\times10^{-3}$ & $4.4$ & $7.66$ & $0.0$ \\
               & $\gamma B_{c2}^\ast(1P)$ & $3.7$ & $1.4$ & $5.52$ & $2.7$ \\[2ex]
$B_{c2}(1D)$ & $\gamma B_{c1}(1P)$ & $58$ & $7$ & $7.25$ & - \\
             & $\gamma B_{c1}(2P)$ & $71$ & $63$ & $116$ & $92.5$ \\
             & $\gamma B_{c2}^\ast(1P)$ & $18$ & $8.8$ & $12.8$ & - \\[2ex]
$B_{c2}(2D)$ & $\gamma B_{c1}(1P)$ & $62$ & $64$ & $112$ & $88.8$ \\
             & $\gamma B_{c1}(2P)$ & $84$ & $15$ & $14.1$ & $0.1$ \\
             & $\gamma B_{c2}^\ast(1P)$ & $19$ & $9.6$ & $27.5$ & $24.7$ \\[2ex]
$B_{c3}(1D)$ & $\gamma B_{c2}^\ast(1P)$ & $149$ & $78$ & $102$ & $98.7$ \\
\end{tabular}
\end{ruledtabular}
\end{table}

The radiative E1 electromagnetic transitions for dominant $D$-wave states are collected in Table~\ref{tab:E1-3}. Again, we compare with some other model predictions~\cite{Godfrey:2004ya, Ebert:2002pp, Eichten:1994gt}. Our results are mostly in accordance with those reported by Ref.~\cite{Eichten:1994gt} and are similar, with some discrepancies, with the results of the remaining references~\cite{Godfrey:2004ya, Ebert:2002pp}. The Table~\ref{tab:E1-3} shows that $D$-wave states are also feasible to measure performing energy scans around their predicted masses when looking at their electromagnetic decay into $P$-wave states, whose masses are almost equally predicted in any theoretical framework mentioned in the Table~\ref{tab:predmassesbc}.

\begin{table}[!t]
\begin{ruledtabular}
\caption{\label{tab:M1-1} The radiative M1 electromagnetic transitions.
We compare with some other model predictions~\cite{Godfrey:2004ya, Ebert:2002pp, Eichten:1994gt}.}
\begin{tabular}{ccrrrr}
Initial & Final & $\Gamma_{\text{The.}}$ & Ref.~\cite{Godfrey:2004ya} & Ref.~\cite{Ebert:2002pp} & Ref.~\cite{Eichten:1994gt} \\
state & state & (eV) & & & \\
\hline
$B_c^\ast(1S)$ & $\gamma B_c(1S)$ & $52$ & $80$ & $33$ & $154.5$\\[2ex]
$B_c^\ast(2S)$ & $\gamma B_c(1S)$ & $650$ & $600$ & $428$ & $123.4$ \\
               & $\gamma B_c(2S)$ & $10$ & $10$ & $17$ & $28.9$ \\[2ex]
$B_c(2S)$ & $\gamma B_c^\ast(1S)$ & $250$ & $300$ & $488$ & $93.3$ \\
\hline
$B_{c2}^\ast(1P)$ & $\gamma B_{c1}(1P)$ & $0.65$ & - & - & - \\
                  & $\gamma B_{c1}(2P)$ & $0.27$ & - & - & - \\[2ex]
$B_{c2}^\ast(2P)$ & $\gamma B_{c1}(1P)$ & $40$ & - & - & - \\
                  & $\gamma B_{c1}(2P)$ & $51$ & - & - & - \\
                  & $\gamma B_{c1}(3P)$ & $0.24$ & - & - & - \\
                  & $\gamma B_{c1}(4P)$ & $0.18$ & - & - & - \\
\hline
$B_{c2}(1D)$ & $\gamma B_{c}^\ast(1D)$ & $2.1\times10^{-6}$ & - & - & - \\[2ex]
$B_{c2}(2D)$ & $\gamma B_{c}^\ast(1D)$ & $0.52$ & - & - & - \\
             & $\gamma B_{c3}^\ast(1D)$ & $1.3\times10^{-4}$ & - & - & - \\[2ex]
$B_{c3}^\ast(1D)$ & $\gamma B_{c2}(1D)$ & $0.20$ & - & - & - \\
\end{tabular}
\end{ruledtabular}
\end{table}

We collect in Table~\ref{tab:M1-1} our predictions for the radiative M1 electromagnetic transitions and compare them with the results of Refs.~\cite{Godfrey:2004ya, Ebert:2002pp, Eichten:1994gt}. Let us give some comments of these results: First, these decay rates are very small, ranging from hundreds to tenths of eV, and even smaller in some cases. Second, the largest rates are found for the radiative M1 electromagnetic transitions between $S$-wave states; although the $B_{c2}^{\ast}(2P)\to \gamma B_{c1}(1P)$ and $B_{c2}^{\ast}(2P)\to \gamma B_{c1}(2P)$ decays have sizeable widths. And third, the theoretical predictions are scarce but, when it is possible to compare, our calculation is in reasonable agreement with those of Refs.~\cite{Godfrey:2004ya, Ebert:2002pp, Eichten:1994gt}.

%%%%%%%%%%%%%%%%%%%%%%%%%%%%%%%%%%%%%%%%%%%%%%%%%%%%%%%%%%%%%%%%%%%%%%%%%%%%%%%%

\begin{table}
\begin{ruledtabular}
\caption{\label{tab:Had-1} Decay rates, in keV, of the spin-nonflip $\pi\pi$ hadronic transitions between spin triplets, and between spin singlets. When possible, we compare with Ref.~\cite{Godfrey:2004ya}.}
\begin{tabular}{rrrr}
Initial & Final & $\Gamma_\text{The.}$ & Ref.~\cite{Godfrey:2004ya} \\
state & state & (keV) & \\
\hline
$2^1S_0$ & $\pi\pi + 1^1S_0$ & $42$ & $57$ \\[2ex]
$2^3S_1$ & $\pi\pi + 1^3S_1$ & $41$ & $57$ \\
\hline
$2^3P_0$ & $\pi\pi + 1^3P_0$ & $12$ & $0.97$ \\
         & $\pi\pi + 1^3P_1$ & $0$ & $0$ \\
         & $\pi\pi + 1^3P_2$ & $5.5\times10^{-3}$ & $5.5\times10^{-2}$ \\[2ex]
$2^3P_1$ & $\pi\pi + 1^3P_0$ & $0$ & $0$ \\
         & $\pi\pi + 1^3P_1$ & $11$ & $2.7$ \\
         & $\pi\pi + 1^3P_2$ & $1.2\times10^{-2}$ & $3.7\times10^{-2}$ \\[2ex]
$2^1P_1$ & $\pi\pi + 1^1P_1$ & $11$ & $1.2$ \\[2ex]
$2^3P_2$ & $\pi\pi + 1^3P_0$ & $1.8\times10^{-2}$ & $1.1\times10^{-2}$ \\
         & $\pi\pi + 1^3P_1$ & $2.0\times10^{-2}$ & $2.1\times10^{-2}$ \\
         & $\pi\pi + 1^3P_2$ & $11$ & $1.0$ \\
\hline
$1^3D_1$ & $\pi\pi + 1^3S_1$ & $0.75$ & $4.3$ \\[2ex]
$1^1D_2$ & $\pi\pi + 1^1S_0$ & $1.1$ & $2.2$ \\[2ex]
$1^3D_2$ & $\pi\pi + 1^3S_1$ & $0.87$ & $2.2$ \\[2ex]
$1^3D_3$ & $\pi\pi + 1^3S_1$ & $0.84$ & $4.3$ \\
\end{tabular}
\end{ruledtabular}
\end{table}

\begin{table}
\begin{ruledtabular}
\caption{\label{tab:Had-2} Other relevant hadronic transitions between $B_c$ states, most of them spin-flip $\pi\pi$ reactions. Decay rates are shown in keV.}
\begin{tabular}{rrr}
Initial & Final & $\Gamma_\text{The.}$ \\
state & state & (keV) \\
\hline
$2^3S_1$ & $\eta + 1^3S_1$ & $0.20$ \\
         & $\pi^0 + 1^1P_1$ & $0.48$ \\
\hline
$1^3P_0$ & $\pi\pi + 1^1S_0$ & $0.58$ \\[2ex]
$2^3P_0$ & $\pi\pi + 1^1S_0$ & $5.5$ \\
         & $\pi\pi + 2^1S_0$ & $7.4\times10^{-2}$ \\[2ex]
$1^3P_1$ & $\pi\pi + 1^1S_0$ & $1.1$ \\[2ex]
$2^3P_1$ & $\pi\pi + 1^1S_0$ & $2.7$ \\
         & $\pi\pi + 2^1S_0$ & $0.15$ \\[2ex]
$1^3P_2$ & $\pi\pi + 1^1S_0$ & $1.6$ \\[2ex]
$2^3P_2$ & $\pi\pi + 1^1S_0$ & $0.85$ \\
         & $\pi\pi + 2^1S_0$ & $2.2\times10^{-2}$ \\
\hline
$1^3D_1$ & $\pi\pi + 1^1P_1$ & $0.13$ \\[2ex]
$1^3D_2$ & $\pi\pi + 1^1P_1$ & $0.17$ \\[2ex]
$1^3D_3$ & $\pi\pi + 1^1P_1$ & $0.16$ \\
\end{tabular}
\end{ruledtabular}
\end{table}

Let us now turn our attention to some, but most relevant, hadronic transitions between $B_c$ mesons. Table~\ref{tab:Had-1} shows our prediction for the decay rates of the spin-nonflip $\pi\pi$ hadronic transitions between spin triplets, and between spin singlets. We compare our results with those reported in Ref.~\cite{Godfrey:2004ya}. In most cases we predict the same order of magnitude, but the diversity of the results makes it difficult to provide general statements. We can mention that the $2^1S_0\to \pi\pi + 1^1S_0$ and $2^3S_1\to \pi\pi + 1^3S_1$ decay rates reported in Ref.~\cite{Godfrey:2004ya} have been fitted following some experimental guidance, whereas they are predictions in our case. In general, our values are larger for spin-nonflip $\pi\pi$ hadronic transitions between $P$-wave states, except in those cases in which the decay width is very small and we predict similar figures. The transitions between $D$-wave states and $S$-wave ones are small due to the only contribution of $C_2$ term in the formulae and our values are slightly smaller than those collected in Ref.~\cite{Godfrey:2004ya}.

From an experimental point of view, independently of the discrepancies between the two theoretical estimations, the $2^1S_0\to \pi\pi + 1^1S_0$ and $2^3S_1\to \pi\pi + 1^3S_1$ transitions have decay rates of about $50\,\text{keV}$ and thus they are potentially observable in experiments. This is in fact the case; however, there is still a lack of statistics which avoids a quantitative study and even to discern if the initial state is either $2^1S_0$ or $2^3S_1$. We find that the $2^3P_J\to \pi\pi+1^3P_J$ transitions have decay rates of the order of $10\,\text{keV}$, making them potentially detectable in experiments. Note that Ref.~\cite{Godfrey:2004ya} predicts an order of magnitude smaller, but what it is clear is that transitions $2^3P_J\to \pi\pi+1^3P_J'$ are very small, with no hope of measuring. And, finally, it seems impossible to explore the $D$-wave states of the $B_c$ system using as an experimental tool the spin-nonflip $\pi\pi$ hadronic transitions.

Table~\ref{tab:Had-2} shows other relevant hadronic transitions between $B_c$ states. Most of them are spin-flip $\pi\pi$ reactions because we are focusing our attention to the $B_c$ mesons which lie below the lowest strong-decay $BD$-threshold and thus there is not enough phase-space to accommodate many light hadrons as part of the final state. As one can see, all decay rates are predicted to be very small with the largest ones being $5.5\,\text{keV}$ and $2.7\,\text{keV}$ for the $2^3P_0\to \pi\pi + 1^1S_0$ and $2^3P_1\to \pi\pi +1^1S_0$ hadronic transition, respectively. It is worth to mention herein that the isospin-violating transition $2^3S_1\to \pi^0+1^1P_1$ has a decay of $0.48\,\text{keV}$, which is of the same order of magnitude than most of the widths collected in Table~\ref{tab:Had-2}; this gives an idea of the smallness of these decay rates. The theoretical computations of these decays are scarce and, if they exist, the way of computing the decay rates results is not very clear and thus we have decided to not collect them in Table~\ref{tab:Had-2}.

%%%%%%%%%%%%%%%%%%%%%%%%%%%%%%%%%%%%%%%%%%%%%%%%%%%%%%%%%%%%%%%%%%%%%%%%%%%%%%%%
%%%%%%%%%%%%%%%%%%%%%%%%%%%%%%%%%%%%%%%%%%%%%%%%%%%%%%%%%%%%%%%%%%%%%%%%%%%%%%%%

\section{Epilogue}
\label{sec:Epilogue}

The properties of the $B_c$-meson family ($c\bar b$) are still not well determined experimentally because the specific mechanisms of formation and decay remain poorly understood. In this article, we have extended our previous investigation of the $B_c$ spectrum to potentially interesting radiative decays and hadronic transitions between $B_c$ states that lie below the lowest strong-decay $BD$-threshold. It is expected that the decay rates of these kind of reactions commensurate the total decay width of such mesons and thus such processes can play an important role on the discovery and quantitative analysis of the $B_c$-meson family.

Our theoretical framework is a non-relativistic constituent quark model in which quark-antiquark and meson-meson degrees of freedom can be incorporated at the same time. Below the $BD$-threshold it is sufficient to work out the naive model which has been widely applied to the charmonium and bottomonium phenomenology, and one expects that it works reasonably well within the $B_c$ sector. The formulae describing radiative E1 and M1 dominant multipole electromagnetic transitions have been used since the early days of heavy quarkonium spectroscopy; we have adapted it to the $c\bar b$ sector and our non-relativistic constituent quark model approach. The calculation of the hadronic decay rates has been performed using the QCDME approach whose unknown constants parametrize the conversion of the emitted gluons into light hadron(s) and have been fitted in previous works. This formalism requires the computation of a hybrid meson spectrum. We have calculated the hybrid states using a natural, parameter-free extension of our quark model based on the Quark Confining String scheme.

Among the results we describe, the following are of particular interest:
\begin{itemize}
\item Below the lowest strong-decay $BD$-threshold, there are two $S$-wave multiplets with spin-parity $0^-$ and $1^-$; another two $P$-wave multiplets with quantum numbers $J^P=0^+$, $1^+$ and $2^+$; and one $D$-wave multiplet with $J^P=1^-$, $2^-$ and $3^-$. Moreover, compared with other theoretical approaches, the predicted spectra are very similar among each other.
\item The radiative E1 electromagnetic transitions between low-lying $B_c$ states present decay rates which range from a few to hundreds of keV. Among the large variety of predictions, it is important to mention that all $S$-, $P$- and $D$-wave states present some electromagnetic decay channels with large widths which would allow their observation, and even their quantitative analysis. Additionally, the radiative M1 electromagnetic transitions are characterized by very small decay rates, ranging from hundreds to tenths of eV; the largest rates are found for the $B_{c2}^{\ast}(2S)\to \gamma B_{c}(1S)$ and $B_{c}(2S)\to \gamma B_{c}^\ast(1S)$ reactions. \\
\item The predicted decay rates of the most relevant hadronic transitions indicate that the spin-nonflip $\pi\pi$ reactions are larger than those where a spin-flip exists. Furthermore, the spin-nonflip $\pi\pi$ hadronic transitions are around $50$, $10$ and $1\,\text{keV}$ between $S$-, $P$- and $D$-wave states, respectively; whereas most of the spin-flip $\pi\pi$ hadronic transitions are of the order of tenths of keV, similar to the case of isospin violating transition $1^3S_1\to \pi^0 + 1^1P_1$.
\end{itemize}

%%%%%%%%%%%%%%%%%%%%%%%%%%%%%%%%%%%%%%%%%%%%%%%%%%%%%%%%%%%%%%%%%%%%%%%%%%%%%%%%
%%%%%%%%%%%%%%%%%%%%%%%%%%%%%%%%%%%%%%%%%%%%%%%%%%%%%%%%%%%%%%%%%%%%%%%%%%%%%%%%

\begin{acknowledgments}
This work has been partially funded by
EU Horizon2020 research and innovation program, STRONG-2020project, under grant agreement no. 824093;
Ministerio Espa\~nol de Ciencia e Innovaci\'on, grant no. PID2019-107844GB-C22 and PID2019-105439GB-C22/AEI/10.13039/501100011033;
and Junta de Andaluc\'ia, contract nos. P18-FR-5057 and Operativo FEDER Andaluc\'ia 2014-2020 UHU-1264517.
\end{acknowledgments}

%%%%%%%%%%%%%%%%%%%%%%%%%%%%%%%%%%%%%%%%%%%%%%%%%%%%%%%%%%%%%%%%%%%%%%%%%%%%%%%%
%%%%%%%%%%%%%%%%%%%%%%%%%%%%%%%%%%%%%%%%%%%%%%%%%%%%%%%%%%%%%%%%%%%%%%%%%%%%%%%%

\bibliography{RadHadDecBc}

\end{document}